\documentclass[twocolumn]{aastex631}
\usepackage{array} 
\usepackage{multirow} 
\usepackage{tikz}
\usepackage{cuted}
\usepackage{amsmath,amssymb}

\usetikzlibrary{positioning, shapes.geometric, arrows.meta}

\hypersetup{linkcolor=red,citecolor=blue,filecolor=red,urlcolor=magenta}
\graphicspath{{fig/}}

\begin{document}
    
    \title{A Markov-Chain-Monte-Carlo-based Hybrid Noise Inference for Continuous Wavelet Power Spectra: with Applications to Solar and Stellar Oscillatory Signals}
    
    \author[0000-0003-4709-7818]{Song Feng}
    \affiliation{Faculty of Information Engineering and Automation, Kunming University of Science and Technology, Kunming 650500, China}
    \email{feng.song@kust.edu.cn (S.F.)}
    
    \author{Lin Li}
    \affiliation{Faculty of Information Engineering and Automation, Kunming University of Science and Technology, Kunming 650500, China}
    
    \author[0000-0002-9514-6402]{Ding Yuan}
    \affiliation{Key Laboratory of Solar Activity and Space Weather, School of Aerospace, Harbin Institute of Technology, Shenzhen, Guangdong 518055, China}
    \email{yuanding@hit.edu.cn (D.Y.)}
    
    \begin{abstract}
        Detecting oscillations in solar and stellar time series is complicated by non-stationary red noise and evolving background emission. Methods based on detrending and AR(1)-based wavelet analysis can introduce spurious periodicities and do not adequately describe time-dependent backgrounds.
        We develop a Bayesian approach that combines the continuous wavelet transform with MCMC sampling to infer a time-dependent background spectrum. The background is represented by a power-law plus white-noise component, with parameters allowed to vary smoothly in time, so that significance levels can be evaluated locally without explicit detrending.
        Tests with synthetic data show that injected oscillations are recovered reliably, while false detections are suppressed in pure-noise cases. Using a frequency-domain signal-to-noise ratio (S/N), we find that oscillations can be identified robustly for $\mathrm{S/N} \gtrsim 2$ under mixed noise conditions. The detectable period range is limited by wavelet resolution, from about 3–4 sampling intervals up to roughly one-quarter of the total duration.
        Application to GOES soft X-ray flare observations shows that the method isolates quasi-periodic oscillations with improved temporal localization compared to standard wavelet and Fourier-based approaches. Meanwhile, this behavior is consistent across a range of noise conditions and signal morphologies.
        
    \end{abstract}
    
    
    \section{Introduction}
    \label{sec:intro}
    
    Oscillations and wave phenomena in the Sun and other stars provide powerful diagnostics of their internal structure, dynamics, and magnetic activity. Through helioseismology and asteroseismology, acoustic (p) and gravity (g) modes enable detailed inference of stellar interiors, constraining key physical parameters such as sound speed, rotation, mixing processes, mass, and age \citep{khomenko2015oscillations, 2016SoPh..291.3143V, 1983Natur.304..689G, 2019LRSP...16....4G}. The study of these oscillations therefore plays an essential role in understanding stellar evolution and magnetohydrodynamic (MHD) processes.
    
    A wide range of spectral and time-frequency analysis techniques have been developed to extract oscillatory components from stellar time series. Commonly used methods include the Fast Fourier Transform (FFT) \citep{2020Ap&SS.365...40L, 2023ApJ...944...16G}, the Continuous Wavelet Transform (CWT) \citep{2018RSPTA.37670253L, 2024ApJ...961..231W, 2018SoPh..293...61D}, Empirical Mode Decomposition \citep{2019MNRAS.488..111D}, and the Synchrosqueezed Transform \citep{2018ApJ...856L..16W}. These approaches \citep{1998BAMS...79...61T, huang1998empirical, daubechies2011synchrosqueezed} provide multi-scale or adaptive representations that are well suited for analyzing non-stationary and intermittent signals.
    
    A central difficulty in oscillation detection arises from the presence of significant background noise. Solar and stellar observations typically contain both red noise—associated with convective motions and magnetic activity—and white noise due to instrumental and photon-counting uncertainties \citep{2015ApJ...798..108I, 2016ApJ...833..284I, 2016ApJ...825..110A}. When oscillation amplitudes are comparable to the noise level, reliable identification becomes challenging, requiring careful statistical modeling of the background.
    
    Two main frameworks are commonly used to address this problem. The first is Bayesian inference in the Fourier domain using Markov Chain Monte Carlo (MCMC), where the power spectrum is modeled as a combination of oscillatory components and stochastic background \citep{2015ApJ...798..108I, 2016ApJ...833..284I, 2020Ap&SS.365...40L, 2023ApJ...944...16G}. This approach provides rigorous uncertainty quantification but typically assumes stationarity and lacks temporal localization. The second approach employs wavelet analysis with a first-order autoregressive (AR(1)) background model for significance testing \citep{1998BAMS...79...61T}. While computationally efficient, the AR(1) model represents a simplified description that may not adequately capture the power-law nature of solar red noise.
    
    Both approaches have important limitations. Fourier-based Bayesian methods are inherently global and cannot resolve the temporal evolution of transient oscillations. Conversely, the conventional wavelet approach may yield biased significance estimates when the true noise deviates from the assumed AR(1) model. Pre-whitening or detrending can partially mitigate this issue \citep{2016ApJ...825..110A}, but such preprocessing may distort the underlying signal and introduce spurious spectral features.
    
    To address these limitations, we develop a hybrid framework that combines the time-frequency localization of the continuous wavelet transform with Bayesian MCMC inference performed directly in the wavelet domain. This approach enables time-dependent estimation of the background spectrum and provides locally adaptive significance thresholds, improving robustness under non-stationary conditions.
    
 The remainder of this paper is organized as follows. Section~\ref{sec:method} describes our methodology. Section~\ref{sec:synthetic} presents validation using synthetic datasets designed to mimic realistic solar variability. Section~\ref{sec:application} applies the proposed method to GOES soft X-ray observations and evaluates its performance under different observational conditions, in comparison with FFT+MCMC and CWT+AR(1) approaches. Section~\ref{sec:discussion} discusses the background noise model and its consistency with observed solar variability. Finally, Section~\ref{sec:conclusion} summarizes the main findings. Details on data availability and reproducibility are provided in Section~\ref{data_avail}.
     
    \begin{figure}[ht!]
        \centering
        \includegraphics[width=0.55\textwidth]{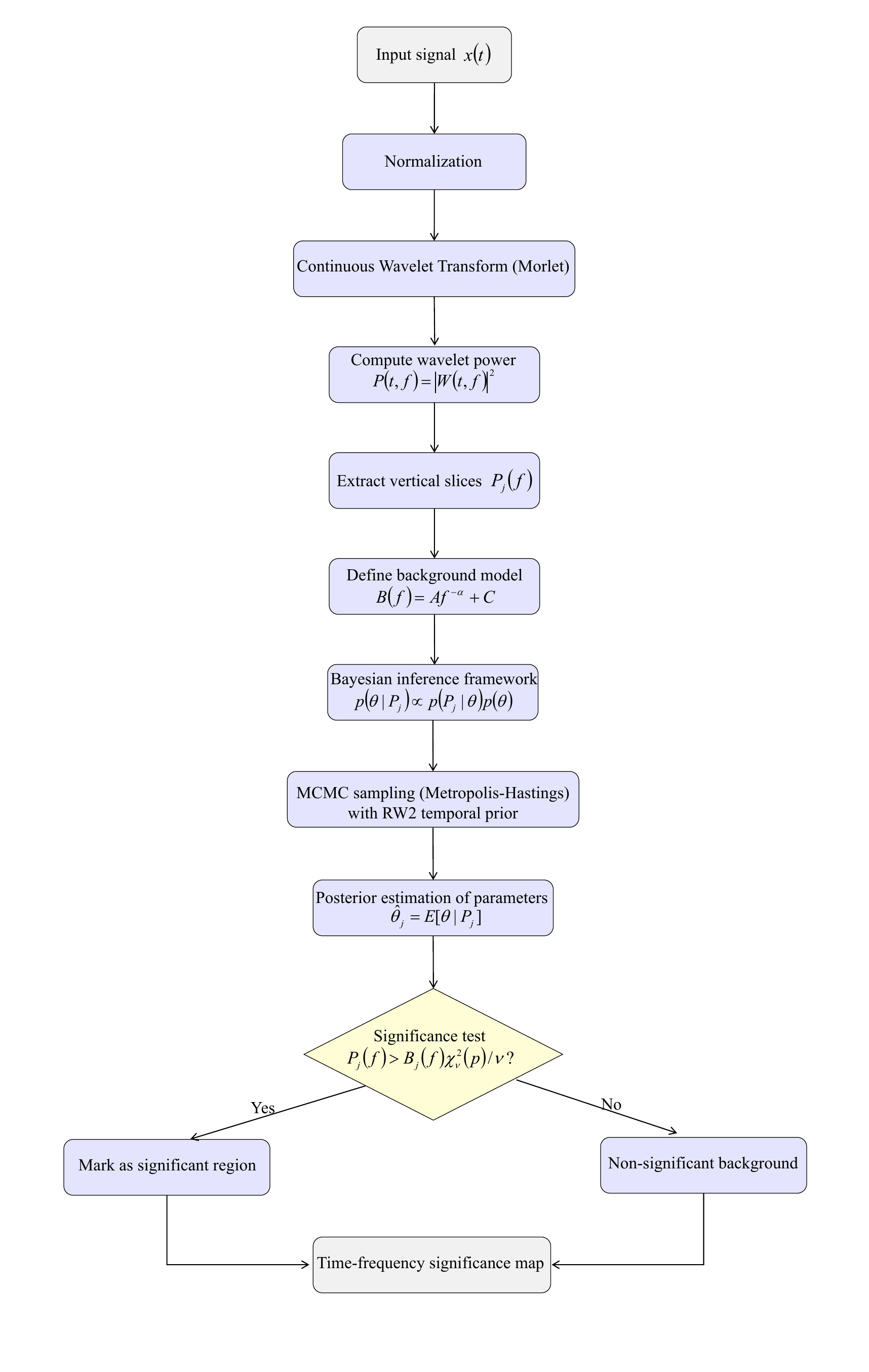}
        \caption{
            Flowchart of the proposed CWT+MCMC framework for wavelet power spectrum estimation and significance testing. 
            The main steps include wavelet transformation, time-dependent background modeling, and Bayesian inference of significance levels.}
        \label{fig:flowchart}
    \end{figure}
    
    \section{Methodology}
    \label{sec:method}
    
An overview of the proposed methodology is illustrated in Figure~\ref{fig:flowchart}. 
        The workflow starts from the input time series $x(t)$, which is first normalized to ensure consistent scaling and numerical stability across different datasets. 
        The normalized signal is then transformed into the time--frequency domain using the Continuous Wavelet Transform (CWT) with a Morlet mother wavelet, producing a complex wavelet coefficient map. 
        From this representation, the wavelet power spectrum is computed as $P(t,f) = |W(t,f)|^2$, providing a localized description of signal power as a function of both time and frequency.
        To enable statistical inference, the two-dimensional power spectrum is decomposed into a sequence of one-dimensional spectra by extracting vertical slices $P_j(f)$ at each time step $t_j$. 
        For each slice, a parametric background model of the form $B(f) = A f^{-\alpha} + C$ is defined, representing a combination of red-noise and white-noise components. 
        Bayesian inference is then performed by constructing the posterior distribution $p(\boldsymbol{\theta}_j \mid P_j) \propto p(P_j \mid \boldsymbol{\theta}_j)p(\boldsymbol{\theta}_j)$, where $\boldsymbol{\theta}_j = \{A_j, \alpha_j, C_j\}$ are the background parameters.
        The posterior distribution is explored using a Metropolis--Hastings MCMC sampler with a second-order random-walk (RW2) temporal prior, which enforces smooth evolution of the parameters across time while suppressing overfitting to local fluctuations. 
        From the resulting posterior samples, parameter estimates are obtained and used to construct the background spectrum at each time step.
        Finally, statistical significance is evaluated by comparing the observed wavelet power with the modeled background. 
        A given time--frequency point is classified as significant if $P_j(f)$ exceeds the corresponding confidence threshold derived from the $\chi^2$ distribution. 
        This decision process produces a binary classification of significant and non-significant regions, which are subsequently combined to form the final time--frequency significance map.
        This pipeline establishes a coherent transformation from the observed time series to statistically interpretable significance estimates, enabling robust detection of oscillatory signals under non-stationary noise conditions without requiring explicit detrending.
    
    \subsection{Continuous Wavelet Transform and Power Spectrum Extraction}
    
    We first normalize the input time series to ensure consistent scaling across different datasets. We then compute the Continuous Wavelet Transform (CWT) using the Morlet wavelet as the mother function, defined as:
    \begin{equation}
        \psi_0(t) = \pi^{-1/4} e^{i\omega_0 t} e^{-t^2/2},
        \label{eq:morlet}
    \end{equation}
    where $\omega_0 = 6$ is the dimensionless frequency parameter satisfying the admissibility condition \citep{1998BAMS...79...61T}.
    
    For a discrete time series $x_n$ with sampling interval $\Delta t$, the CWT is given by:
    \begin{equation}
        W(n, s) = \sqrt{\frac{\Delta t}{s}} \sum_{n'} x_{n'} \psi_0^* \left[ (n' - n) \frac{\Delta t}{s} \right],
        \label{eq:cwt}
    \end{equation}
    where $n$ is the discrete time index corresponding to $t = n \Delta t$. $s$ represents the wavelet scale, and $*$ denotes complex conjugation. The corresponding wavelet power spectrum is defined as:
    \begin{equation}
        P(t, s) = |W(t, s)|^2.
        \label{eq:power}
    \end{equation}
    
    We convert scales to frequencies using $f = (\omega_0 + \sqrt{2 + \omega_0^2})/(4\pi s)$, yielding the frequency-dependent representation $P(t, f)$. For subsequent analysis, we extract temporal slices $P_j(f) = P(t_j, f)$ at each time $t_j$, which represent local spectral power distributions.
    
    \subsection{Bayesian Background Noise Modeling}
    
    For each temporal slice $P_j(f)$, we estimate the parameters of the background noise model within a Bayesian framework using MCMC sampling. The posterior distribution of parameters $\boldsymbol{\theta}_j$ is expressed as:
    \begin{equation}
        p(\boldsymbol{\theta}_j | P_j) \propto \mathcal{L}(P_j | \boldsymbol{\theta}_j) \, p(\boldsymbol{\theta}_j),
        \label{eq:posterior}
    \end{equation}
    where $\mathcal{L}$ denotes the likelihood and $p(\boldsymbol{\theta}_j)$ the prior.
    The background power spectrum at time $t_j$ is modeled as:
    \begin{equation}
        B(f; \boldsymbol{\theta}_j) = A_j f^{-\alpha_j} + C_j,
        \label{eq:background}
    \end{equation}
    where $\boldsymbol{\theta}_j = \{A_j, \alpha_j, C_j\}$ correspond to the power-law amplitude, spectral index, and white-noise level, respectively.
    
    Assuming a locally stationary background, the wavelet power at each frequency approximately follows a $\chi^2$ distribution with $\nu = 2$ degrees of freedom for the Morlet wavelet. This approximation follows the standard assumption for wavelet power spectra under locally stationary processes \citep{1998BAMS...79...61T}. The likelihood is therefore:
    \begin{equation}
        \mathcal{L}(P_j | \boldsymbol{\theta}_j)
        = \prod_f \frac{1}{B(f; \boldsymbol{\theta}_j)}
        \exp\left[-\frac{P_j(f)}{B(f; \boldsymbol{\theta}_j)}\right].
        \label{eq:likelihood}
    \end{equation}
    
    Although Equation~(\ref{eq:likelihood}) assumes independence across frequencies, neighboring wavelet coefficients exhibit finite correlation due to the bandwidth of the Morlet wavelet. In practice, this approximation is sufficient because inference targets smoothly varying background parameters rather than individual spectral bins, and temporal regularization suppresses overfitting to correlated fluctuations.
    
    \subsection{Prior Distributions and Temporal Smoothing}
    
    We assign weakly informative priors based on physical considerations:
    \begin{align}
        A_j &\sim \mathcal{N}^+(\mu_A, \sigma_A^2), \\
        \alpha_j &\sim \mathcal{N}(\mu_\alpha, \sigma_\alpha^2), \\
        C_j &\sim \mathcal{N}^+(\mu_C, \sigma_C^2),
    \end{align}
    where $\mathcal{N}^+$ denotes a half-normal distribution enforcing positivity.
    
The adopted power-law plus constant formulation is motivated by observational evidence that solar and stellar background fluctuations exhibit red-noise behavior with power spectral densities of the form $S(f) \propto f^{-\beta}$, typically with $\beta \sim 1.5--4.5$  \citep[e.g.,][]{2015ApJ...798..108I, 2016ApJ...833..284I, 2016ApJ...825..110A, 2018SoPh..293...61D, 2020Ap&SS.365...40L, 2023ApJ...944...16G}. The additive constant term accounts for high-frequency white-noise contributions arising from photon-counting statistics and instrumental effects.
    
    To capture temporal evolution while avoiding overfitting, we impose a second-order random-walk (RW2) prior:
    \begin{equation}
        \theta_{j} - 2\theta_{j-1} + \theta_{j-2} \sim \mathcal{N}(0, \tau^{-1}),
        \label{eq:rw2}
    \end{equation}
    which penalizes rapid curvature in parameter trajectories, thereby enforcing smooth temporal evolution of the background parameters. The precision parameter $\tau$ controls the degree of smoothness and is assigned a weakly informative hyperprior. Empirically, posterior estimates are stable across a broad range of $\tau$ values.
    
    Within a finite frequency range, partial degeneracy may arise between $A_j$ and $\alpha_j$, while $C_j$ can overlap with the high-frequency tail of the power-law component. These effects are mitigated through joint inference across frequencies, weakly informative priors, and temporal coupling enforced by the RW2 prior.
    
    \subsection{MCMC Sampling and Implementation}
    
    Posterior inference is performed using a Metropolis--Hastings MCMC sampler with adaptive proposal distributions. Each chain is run for $2 \times 10^4$ iterations, with the first 25\% discarded as burn-in and thinning applied to reduce autocorrelation. 
    Convergence is assessed using standard diagnostics, including trace inspection, effective sample size, and the Gelman--Rubin statistic ($\hat{R} < 1.05$).
    
    For statistical significance testing, we compute local $p$-level thresholds:
    \begin{equation}
        T_j(f) = B_j(f; \boldsymbol{\theta}_j) \cdot \chi^2_{\nu}(p)/\nu, \quad \nu = 2,
        \label{eq:threshold}
    \end{equation}
    where $\chi^2_{\nu}(p)$ denotes the $p$-th quantile of the chi-squared distribution.
    
    \subsection{Comparison Methods}
    
    Throughout this study, we compare three representative analysis approaches: (1) FFT + MCMC \citep{2023ApJ...944...16G}: Bayesian modeling of the Fourier power spectrum using the same power-law background model as in Equation~(\ref{eq:background}), providing global characterization of the noise properties. (2) CWT + AR(1) \citep{1998BAMS...79...61T}: Conventional wavelet analysis using a stationary AR(1) noise model for significance testing. It offers time-frequency localization but assumes stationary background statistics. (3) CWT + MCMC (proposed): The method developed in this work, combining wavelet time-frequency localization with Bayesian inference of a time-evolving background spectrum, enabling adaptive significance testing under non-stationary conditions.
    All methods are applied consistently to both synthetic and observational datasets to ensure a fair and controlled comparison.

\begin{figure*}[htbp]
    \centering
    \includegraphics[width=0.95\textwidth]{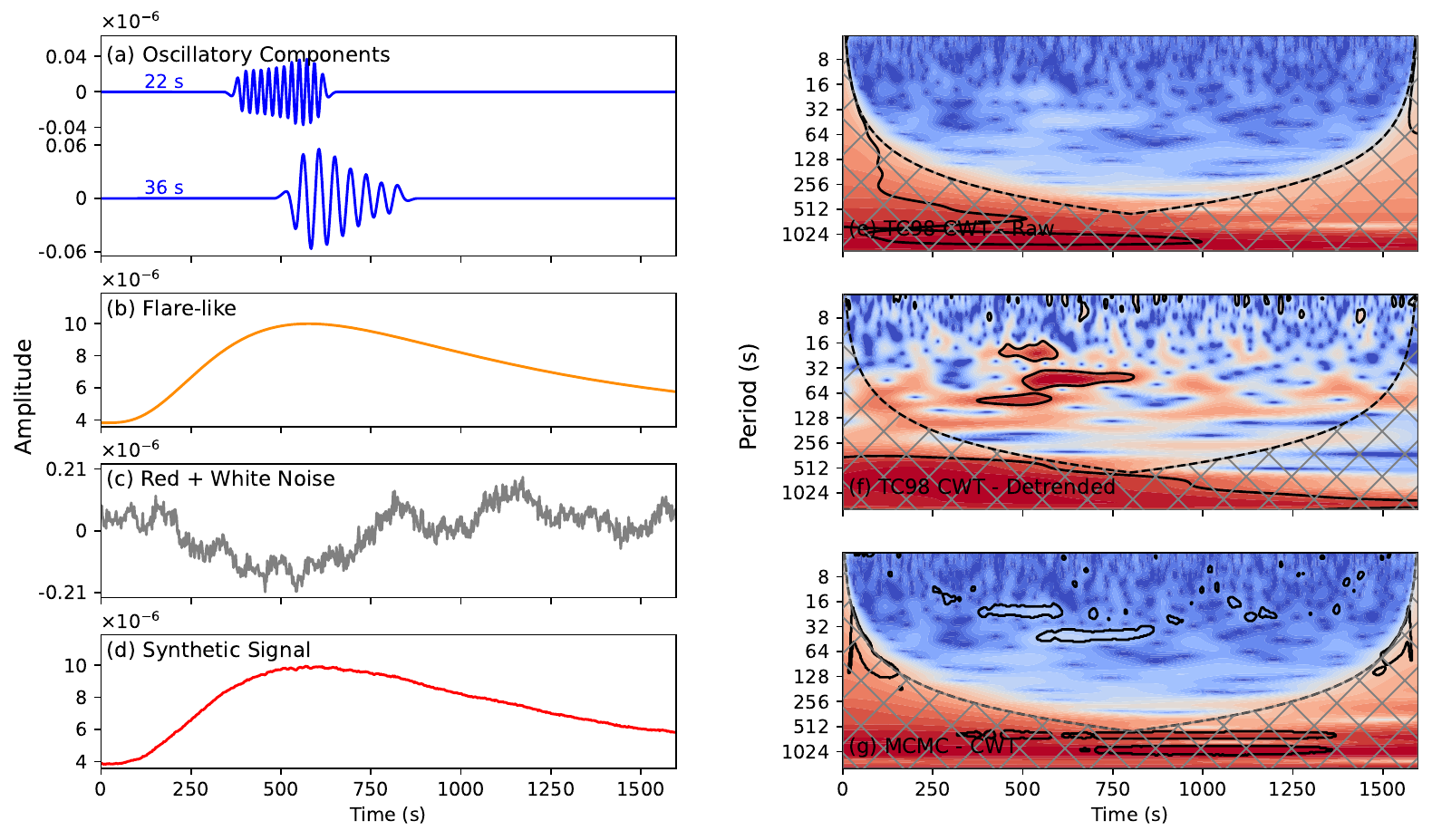} 
    \caption{
        Synthetic test with a simple signal. 
        (a)--(d) Individual components: oscillatory signal,  flare background, red + white noise. 
        (e) Wavelet power spectrum from the standard CWT+AR(1) method. 
        (f) Result after detrending (80\,s) followed by CWT+AR(1) analysis. 
        (g) Result from the CWT+MCMC method. 
        Black contours indicate the 95\% confidence level.
    }
    \label{fig:synthetic_signal}
\end{figure*}

\begin{figure*}[htbp]
    \centering
    \includegraphics[width=0.98\textwidth]{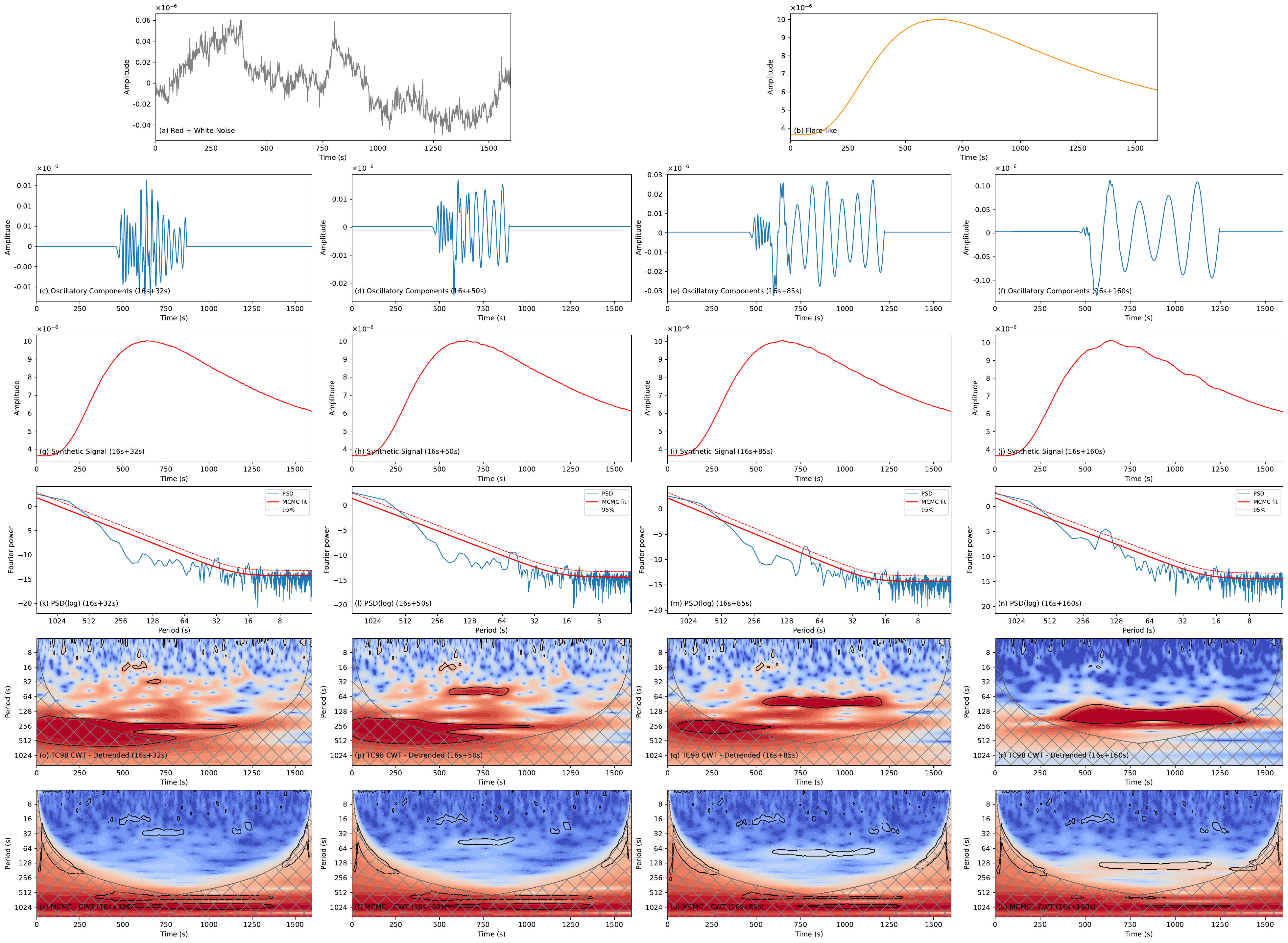}
    \caption{
        Synthetic signals with multiple non-stationary oscillatory components. 
        Each column corresponds to an independent realization with the same background but different oscillation periods. 
        From top to bottom: 
        (1) background components (red + white noise with flare profile); 
        (2) injected oscillations with time-varying amplitudes and periods; 
        (3) composite signals; 
        (4) global Fourier power spectra (FFT+MCMC); 
        (5) detrended wavelet spectra (CWT+AR(1), detrending scale 290\,s); 
        (6) wavelet spectra from the CWT+MCMC method. 
        One component has a characteristic period of $\sim$16\,s, while the second varies across columns (32\,s, 50\,s, 85\,s, and 160\,s). 
        Black contours mark the 95\% confidence level.
    }
    \label{fig:complex_synthetic}
\end{figure*}
    
 \section{Synthetic Tests}
 \label{sec:synthetic}
 
 \subsection{Synthetic Data Construction}
 
 Synthetic time series are constructed to reproduce key features of solar flare observations, including a non-stationary background, quasi-periodic oscillations, and mixed noise. The signal is written as
 \begin{equation}
     s(t) = s_{\mathrm{flare}}(t) + s_{\mathrm{osc}}(t) + n_{\mathrm{red}}(t) + n_{\mathrm{white}}(t).
 \end{equation}
 
 The flare component follows an asymmetric rise–decay profile,
 \begin{equation}
     \mathrm{flare}(t)=
     \begin{cases}
         0, & t<t_0,\\
         A\left(1-e^{-\frac{t-t_0}{\tau_{\mathrm{rise}}}}\right)
         e^{-\frac{t-t_0}{\tau_{\mathrm{decay}}}}, & t\ge t_0,
     \end{cases}
 \end{equation}
 which approximates the temporal evolution of GOES soft X-ray light curves.  
 
 The oscillatory component consists of two modes,
 \begin{equation}
     s_{\mathrm{osc}}(t)
     = A_1(t)\cos[2\pi \phi_1(t)]
     + A_2(t)\cos[2\pi \phi_2(t)],
 \end{equation}
 with slowly varying amplitudes and periods. The noise background includes a red-noise component with $S(f)\propto f^{-\beta}$ ($\beta \approx 2$) and an additive white-noise term. The cadence is $\Delta t = 2$~s.

 \begin{figure*}[htbp]
     \centering
     \includegraphics[width=0.95\textwidth]{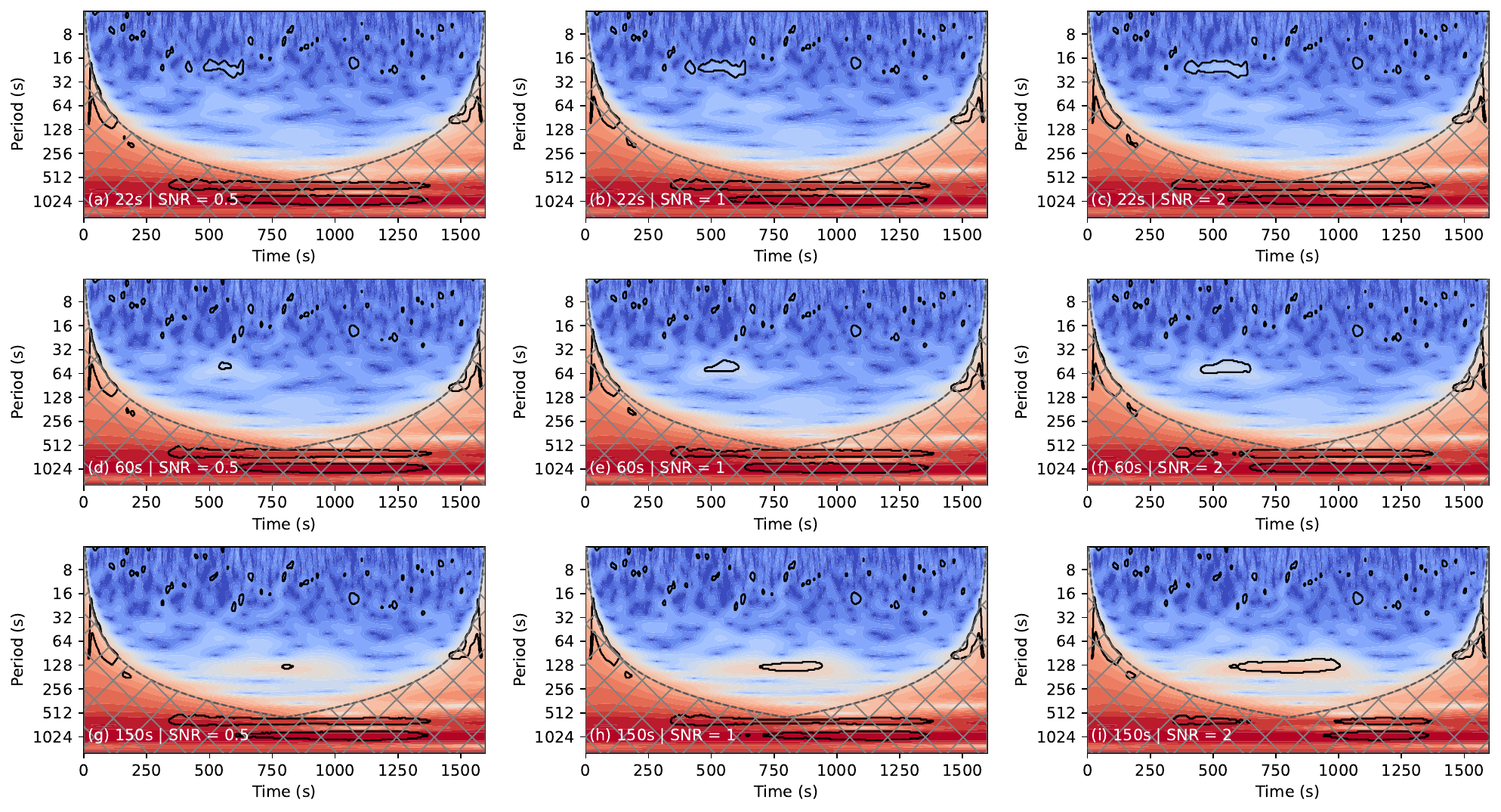}
     \caption{Detection results for different signal-to-noise ratios (SNRs) and oscillation periods.  Three representative periods are considered: 22\,s (top), 60\,s (middle), and 150\,s (bottom).  Columns correspond to $\mathrm{SNR}=0.5$, 1, and 2.  Black contours indicate regions above the 95\% confidence level.
     }
     \label{fig:snr_multiband}
 \end{figure*}
 
 \subsection{Detection Performance and Limitations}
 
Figure~\ref{fig:synthetic_signal} shows a simple realization consisting of two oscillatory components embedded in a flare-like background with red and white noise. In the CWT+AR(1) result without detrending (panel e), no regions exceed the 95\% confidence level, and the oscillatory components are not identified. After detrending with a cutoff of 80\,s (panel f), the two oscillatory components become partially visible; however, an additional feature appears near the detrending scale. This feature does not correspond to any injected signal and reflects the influence of the filtering process. In addition, the temporal extent of the two oscillatory components is only partially recovered. In contrast, the CWT+MCMC result (panel g) identifies both oscillatory components over their full duration, without introducing additional features.
 
Figure~\ref{fig:complex_synthetic} extends this comparison to signals with multiple non-stationary components. Each realization contains two oscillations with time-dependent amplitudes and periods, superimposed on the same background. The Fourier spectra (row 4) show peaks corresponding to the dominant periods but provide no information on their temporal evolution. The detrended wavelet results (row 5) recover parts of the oscillatory components, but the detected regions are often fragmented, particularly when the periods vary in time or approach the detrending scale. In contrast, the CWT+MCMC results (row 6) show continuous structures in the time--frequency plane that follow the evolution of both components across all realizations, including cases with longer periods.
 
The detectable period range is limited by wavelet resolution and edge effects. For a time series of duration $T$, reliable detections are typically confined to periods between $\sim 3$--$4\,\Delta t$ and $\sim T/4$.

Figure~\ref{fig:snr_multiband} shows the dependence on signal-to-noise ratio (SNR) for three representative periods. At $\mathrm{SNR}=0.5$, the detected power is weak and does not form coherent structures. At $\mathrm{SNR}=1$, localized regions begin to appear but remain discontinuous, especially at shorter periods where the noise contribution is stronger. For $\mathrm{SNR}=2$, the oscillatory components are recovered as continuous regions across all three periods. This trend is consistent for short-, intermediate-, and long-period signals, indicating that the detectability depends on both the signal amplitude and the local background level. 
 
\textbf{Additional tests with different noise realizations show similar behavior, with only minor variations across different noise levels and spectral slopes. We note that the SNR adopted here is defined in the wavelet domain as the ratio between the peak oscillatory power and the local background power, rather than the conventional amplitude-based observational SNR. Therefore, the selected range ($\mathrm{SNR}=0.5$--2) represents weak-to-moderate oscillatory signals relative to the local background and is intended to mimic challenging observational conditions where signal detection becomes difficult. The threshold identified here is empirical rather than strict, and may vary depending on the noise structure and temporal localization of the signal.}

\section{Application to GOES Soft X-ray Observations}
\label{sec:application}

\subsection{Data and Comparative Methods}

To evaluate performance on observational data, we analyze GOES 1--8~\AA{} soft X-ray light curves under three representative scenarios: (1) a quiet-Sun interval without flare activity, (2) a flare event exhibiting quasi-periodic pulsations (QPPs), and (3) a flare event without identifiable QPP signatures.
These scenarios allow us to assess detection capability, false-positive behavior, and robustness under realistic solar conditions.
We selected observational data of flares with QPPs and without QPPs during their eruption, following the work of \citet{2023ApJ...944...16G}.
The three methods described in Section~\ref{sec:method} (FFT+MCMC, CWT+AR(1), and the proposed CWT+MCMC) are applied consistently to all datasets.

\begin{figure*}[htbp]
    \centering
    \includegraphics[width=0.95\textwidth]{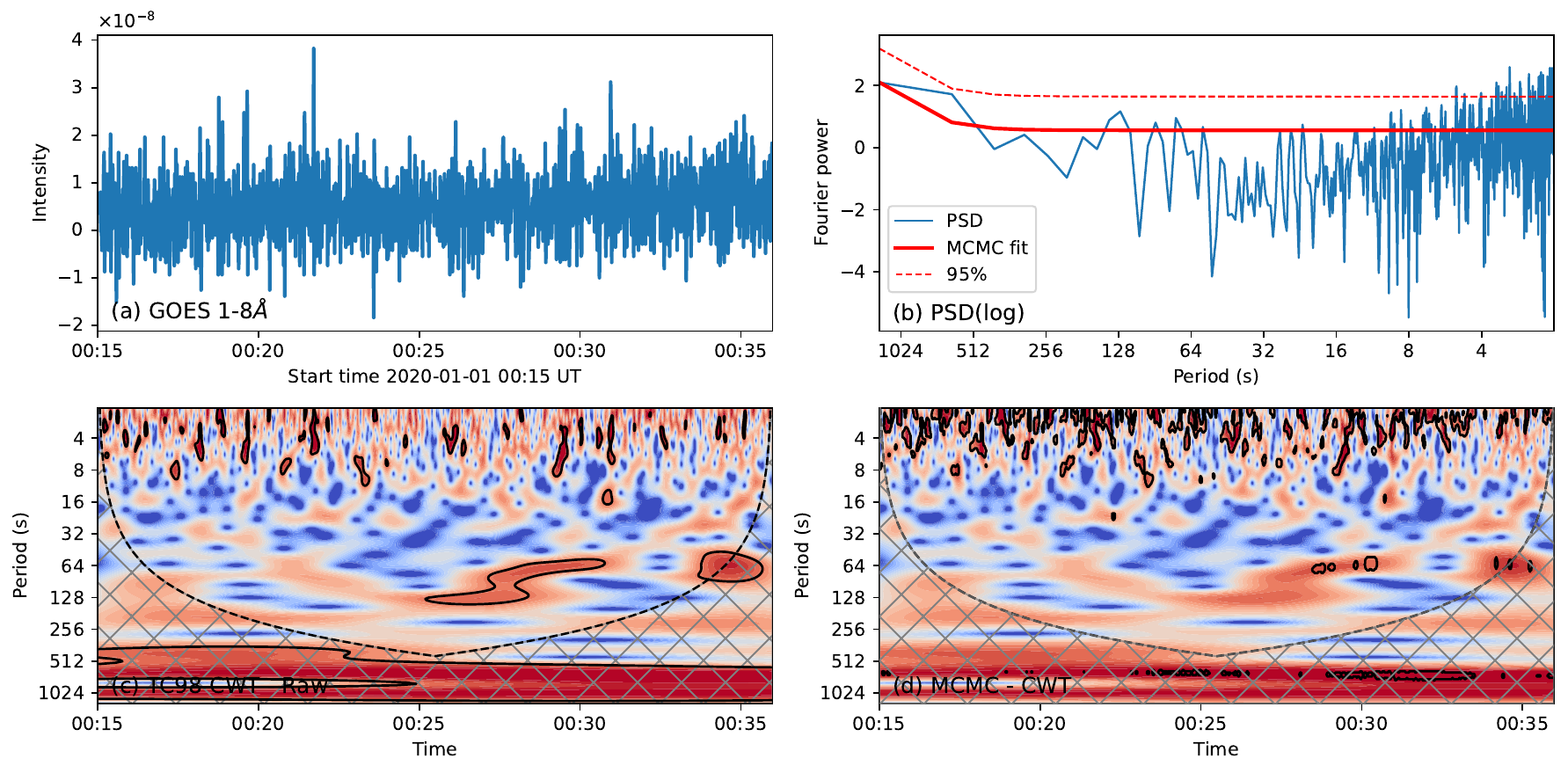}
    \caption{
        GOES 1--8~\AA{} light curve for a quiet-Sun interval. 
        (a) Observed soft X-ray flux. 
        (b) Fourier power spectrum (FFT+MCMC). 
        (c) Wavelet power spectrum from the CWT+AR(1) method. 
        (d) Wavelet power spectrum from the CWT+MCMC method. 
        A localized feature near $\sim$128~s is present in panel (c) but not in the other representations. 
        Black contours indicate the 95\% confidence level.
    }
    \label{fig:goes_quietSun}
\end{figure*}

\subsection{Quiet-Sun Interval (No Oscillation)}
Figure~\ref{fig:goes_quietSun} shows the analysis of a flare-free quiet-Sun interval, which serves as a benchmark for evaluating false-positive behavior under noise-dominated conditions.
    The light curve exhibits only low-amplitude variability, with no visually identifiable oscillatory signatures, and is therefore expected to contain no significant periodic components.
    The FFT+MCMC method yields a smooth power spectrum without distinct peaks, consistent with the absence of coherent oscillations.
    In contrast, the CWT+AR(1) method identifies a localized region of apparent significance around $\sim$128~s.
    This feature is not supported by the Fourier analysis and is most likely a false-positive detection arising from the mismatch between the assumed AR(1) background model and the true power-law-like noise structure of the data.
    Such mismatch leads to underestimated background levels at low frequencies, artificially enhancing apparent significance.
    The proposed CWT+MCMC method does not detect any statistically significant oscillatory features across the entire time--frequency domain.
    This behavior indicates that the adaptive background estimation successfully tracks the underlying noise structure and prevents spurious detections.
    The result demonstrates that the method maintains reliable false-positive control under realistic, non-stationary noise conditions, which is essential for robust oscillation analysis.

\begin{figure*}[htbp]
    \centering
    \includegraphics[width=0.90\textwidth]{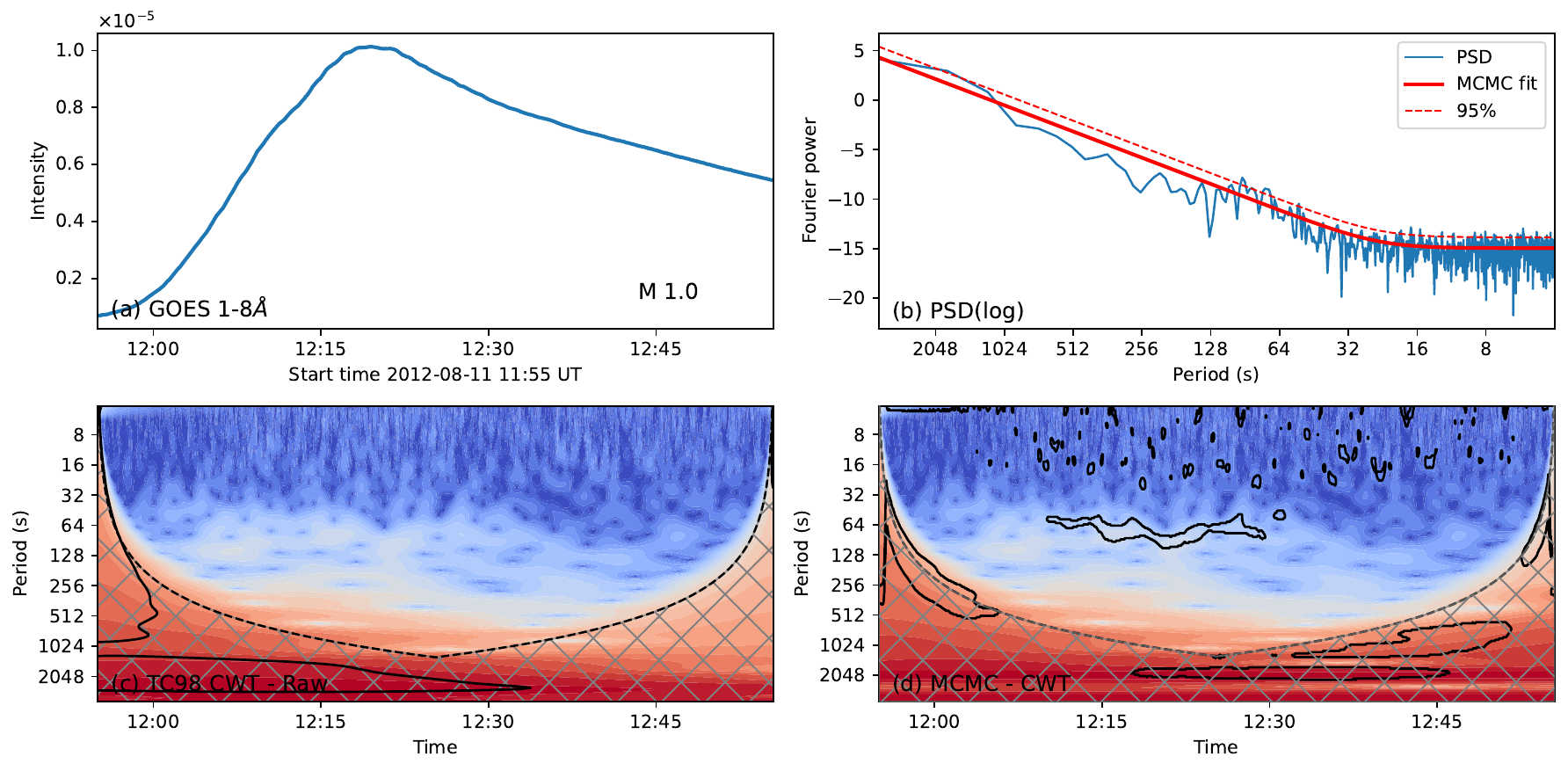}
    \caption{
        GOES 1--8~\AA{} light curve for a flare with quasi-periodic pulsations. 
        (a) Observed soft X-ray flux. 
        (b) Fourier power spectrum (FFT+MCMC). 
        (c) Wavelet power spectrum from the CWT+AR(1) method. 
        (d) Wavelet power spectrum from the CWT+MCMC method. 
        Enhanced power near $\sim$80~s is visible in the Fourier spectrum, and a localized component at a similar period appears in the time--frequency domain. 
        Black contours indicate the 95\% confidence level.
    }
    \label{fig:flare_qpp}
\end{figure*}

\subsection{Flare with QPP Signatures}
Figure~\ref{fig:flare_qpp} presents the analysis of a flare event exhibiting quasi-periodic pulsations (QPPs), providing a test case for detection sensitivity under strongly non-stationary conditions.
The light curve shows pronounced variability associated with the flare evolution, but no clear oscillatory signatures can be directly identified by visual inspection.
Any potential periodic components are therefore likely to be weak and embedded within the rapidly evolving background, making their detection non-trivial without a dedicated time--frequency analysis framework.
The FFT+MCMC method (Figure~\ref{fig:flare_qpp}b) reveals enhanced spectral power near $\sim$80~s, indicating the presence of a dominant periodic component.
However, due to the global nature of the Fourier transform, it is not possible to determine when this oscillation occurs or how it evolves during different phases of the flare.
In contrast, the CWT+MCMC method (Figure~\ref{fig:flare_qpp}) identifies a statistically significant and temporally localized oscillatory component near $\sim$80~s during the impulsive phase of the flare, but the CWT+AR(1) method fails to identify it.
The adaptive background model captures the evolving noise conditions, allowing the significance threshold to adjust accordingly.
\textbf{Unlike the CWT+AR(1) method, where significance regions closely follow strong wavelet power because of the fixed background assumption, the confidence regions in the CWT+MCMC method are determined jointly by the local wavelet power and the posterior uncertainty of the inferred background model.
Therefore, regions of enhanced power do not necessarily correspond directly to high-confidence regions if the local background level or its uncertainty is also elevated.}
This leads to improved sensitivity while maintaining statistical consistency, demonstrating the advantage of the proposed method in detecting transient oscillations embedded in non-stationary backgrounds.

\subsection{Flare without QPP Signatures}
\begin{figure*}[htbp]
    \centering
    \includegraphics[width=0.90\textwidth]{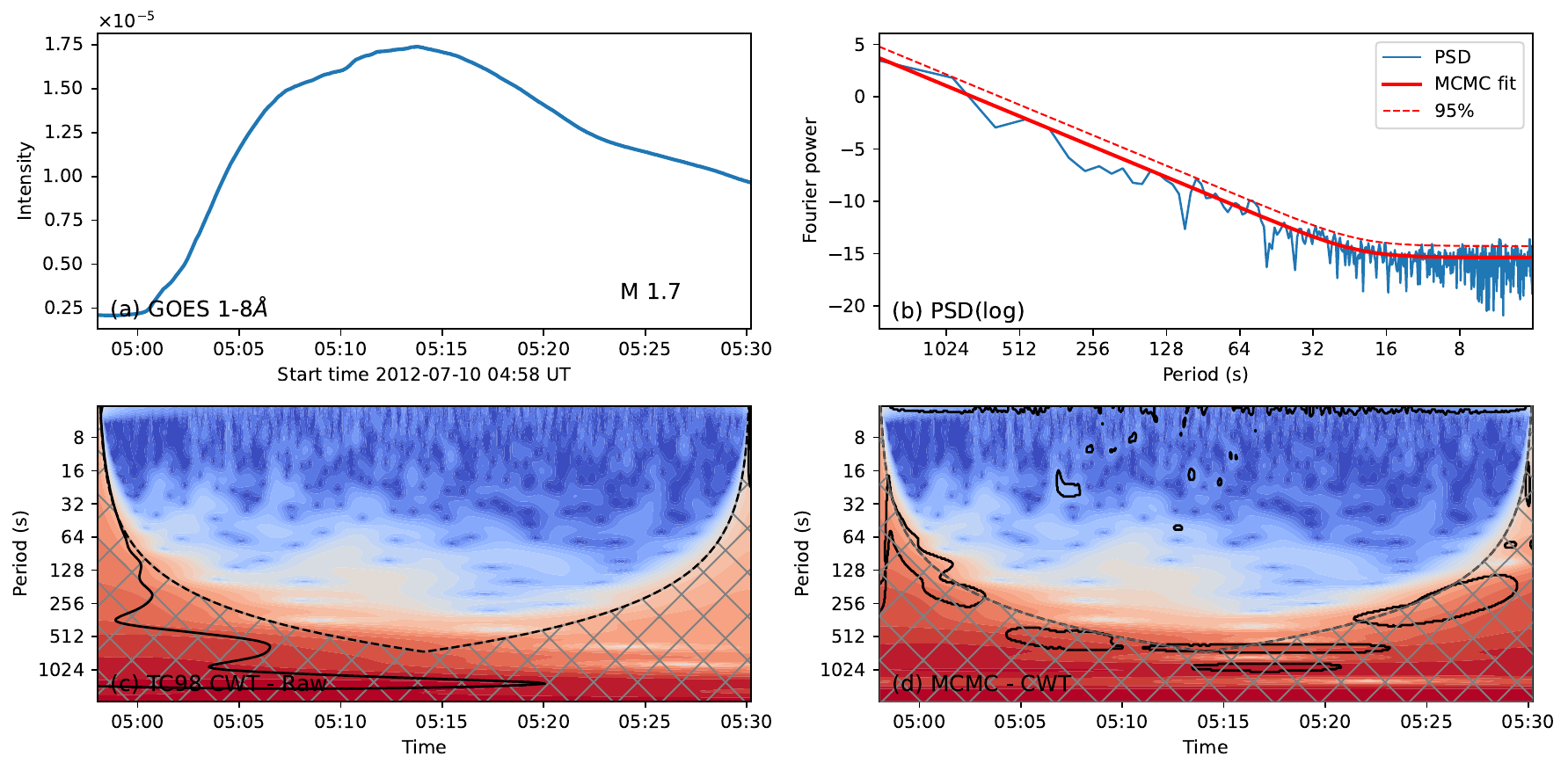}
    \caption{Detection results for a GOES soft X-ray flare without evident quasi-periodic pulsations (QPPs).   (a) Observed 1--8~\AA{} light curve showing a typical flare profile without clear oscillatory signatures.  (b) FFT+MCMC: the Fourier spectrum does not exhibit distinct periodic components.  (c) CWT+AR(1): no statistically significant oscillatory power is detected, indicating that the method does not produce false positives in this case.   (d) CWT+MCMC (this work): no statistically significant oscillatory power is detected, demonstrating robustness under non-stationary flare conditions. The 95\% confidence level is indicated by contour lines.}    
    \label{fig:flare_noqpp}
\end{figure*}

Figure~\ref{fig:flare_noqpp} shows a flare event without clear quasi-periodic pulsations, representing a challenging scenario where strong non-stationary background variations are present but no true oscillatory signal exists.
    The light curve is dominated by the flare evolution, which introduces substantial low-frequency power and time-varying background levels.
    The FFT+MCMC analysis does not reveal any distinct periodic components, indicating that no dominant oscillation is present in the global spectrum.
    The CWT+AR(1) method similarly does not identify statistically significant oscillatory power, suggesting that, in this case, the background model does not produce obvious false positives.
    The proposed CWT+MCMC method also yields no significant detections across the time--frequency domain.
    Importantly, despite the strongly non-stationary background associated with the flare, the method does not introduce spurious oscillatory features.
    This result confirms that the adaptive background modeling and Bayesian inference framework effectively distinguishes between genuine oscillatory signals and complex, non-periodic variability.
    The consistency of results across all three methods supports the conclusion that the observed variability is dominated by non-periodic flare dynamics rather than oscillatory processes.
    More importantly, it demonstrates that the proposed method maintains both sensitivity and specificity, achieving reliable detection when oscillations are present and robust suppression of false positives when they are absent.

\subsection{Comparative Performance}
Across the three observational scenarios, the three methods exhibit systematic and interpretable differences that reflect their underlying assumptions and modeling capabilities.
    The FFT+MCMC approach effectively characterizes the global spectral properties of the signal and can identify dominant periodic components when present.
    However, due to its inherently global formulation, it lacks temporal resolution and cannot determine when an oscillation occurs or how it evolves during the flare.
    The CWT+AR(1) method provides time--frequency localization, enabling the identification of transient features.
    Nevertheless, its reliance on a stationary AR(1) background model limits its applicability under realistic solar conditions, where the noise properties evolve significantly over time.
    As a result, the method may either produce spurious detections, as seen in the quiet-Sun case, or miss genuine oscillatory signals when the background is strongly non-stationary, as in the QPP flare scenario.
    In contrast, the proposed CWT+MCMC method combines time--frequency localization with a time-dependent statistical model of the background.
    By explicitly modeling the evolution of the noise spectrum, the method adapts its significance thresholds to local conditions in both time and frequency.
    This enables consistent performance across all scenarios: it avoids false positives in noise-dominated data, successfully detects and localizes oscillatory signals when present, and remains robust under strongly non-stationary flare conditions.
    Taken together, these results demonstrate that the proposed method achieves a balance between sensitivity and specificity, providing reliable oscillation detection while maintaining statistical consistency across diverse observational regimes.

\section{Discussion}
\label{sec:discussion}

\subsection{Background Model and Observational Consistency}

To examine whether the adopted background model is consistent with observations, we compare it with power spectral densities derived from GOES 1--8~\AA{} light curves. Fourier spectra are computed for representative intervals, including both flare and pre-flare phases, and fitted with the form $P(f) = A f^{-\alpha} + C$.

\begin{figure*}[htbp]
    \centering
    \includegraphics[width=0.90\textwidth]{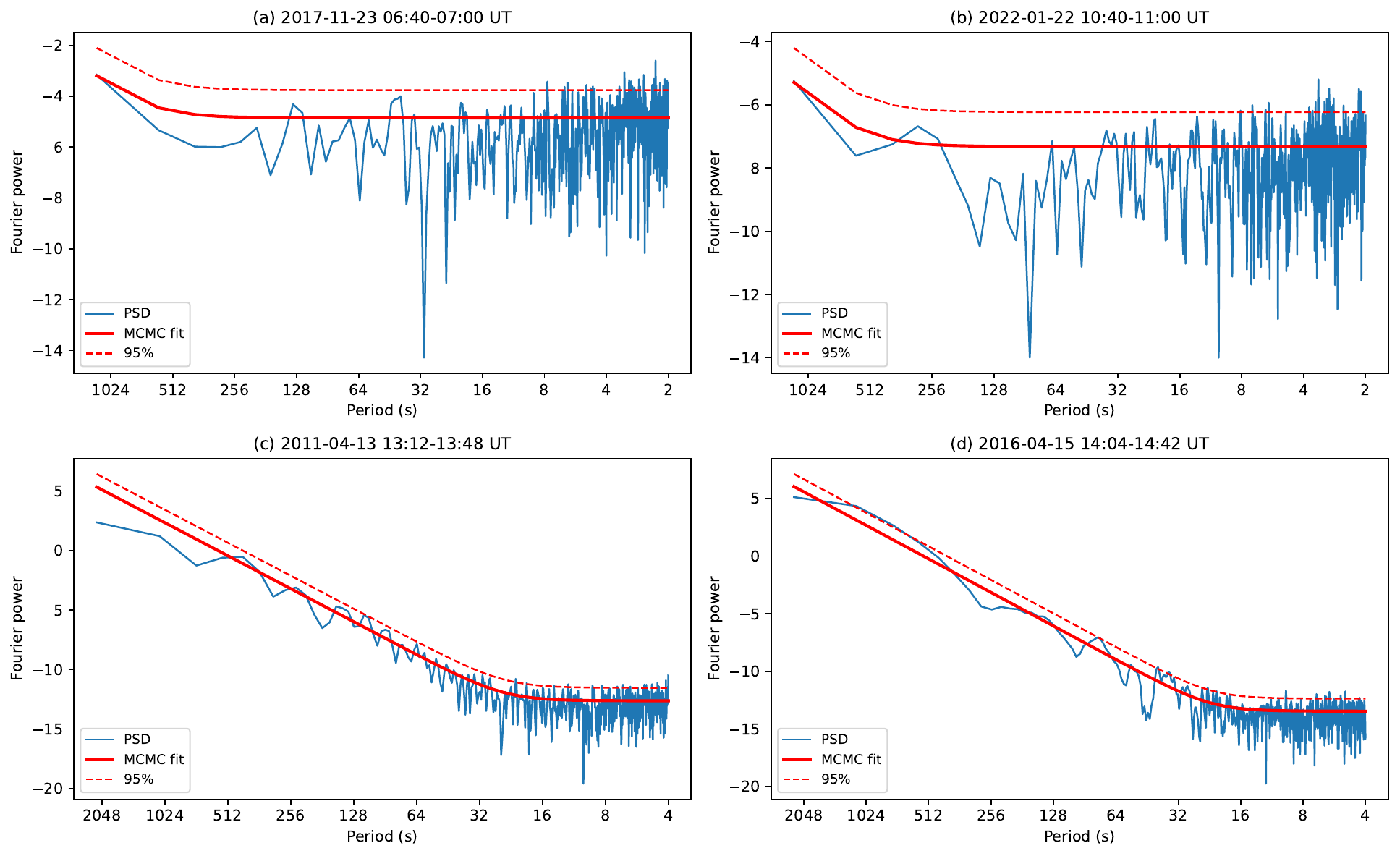}
    \caption{
        Fourier power spectra from representative GOES 1--8~\AA{} time series segments. 
        The spectra are fitted with the model $P(f)=A f^{-\alpha}+C$. 
        A power-law trend is present at low frequencies, with a gradual flattening at higher frequencies. 
        The fitted spectral indices lie in the range $\alpha \sim 2.5$--4.5. }     \label{fig:noise_model_validation}
\end{figure*}

As shown in Figure~\ref{fig:noise_model_validation}, the spectra follow a power-law distribution over a broad frequency range. The fitted indices typically fall between $\alpha \sim 2.5$ and $4.5$, with a gradual flattening at higher frequencies. This behavior is consistent across different intervals and reflects the transition from red-noise-dominated to white-noise-dominated regimes.

The same functional form is used in the time-dependent background model (Section~\ref{sec:method}), and the agreement in Figure~\ref{fig:noise_model_validation} indicates that it provides a reasonable description of the observed variability over the frequency range relevant for wavelet analysis.

\subsection{Implications for Oscillation Detection}

The results highlight several points relevant to oscillation analysis in non-stationary time series. 

First, preprocessing steps such as detrending or high-pass filtering modify the power spectrum in a frequency-dependent way. In particular, applying a filter with transfer function $H(f)$ leads to $S_{\tilde{x}}(f) = |H(f)|^2 S_x(f)$, which alters both the signal and the background near the cutoff scale. This effect can introduce additional structure in the spectrum and complicate the interpretation of significance levels.

Second, the choice of background model affects the resulting detection thresholds. The commonly used AR(1) formulation assumes an exponential spectrum, whereas observed data are better described by a power-law form. Differences between these representations are most pronounced at low frequencies, where the background level contributes directly to significance estimates.

Third, Fourier-based methods and wavelet-based methods emphasize different aspects of the signal. Fourier spectra represent global power distributions, while wavelet spectra resolve temporal evolution. In non-stationary conditions, localized structures in the time--frequency domain may not appear as distinct peaks in the global spectrum.

The CWT+MCMC approach combines these aspects by estimating a time-dependent background and evaluating significance locally in both time and frequency. In the examples considered here, this leads to stable behavior across different noise conditions and variability levels, without requiring explicit detrending.

\section{Conclusion}
\label{sec:conclusion}

We have developed a Bayesian framework for oscillation detection in non-stationary solar and stellar time series by combining the continuous wavelet transform with time-dependent background estimation. The background spectrum is described by a power-law plus constant component, with parameters allowed to vary smoothly in time.

Tests with synthetic data show that oscillatory signals can be recovered over a range of periods and noise conditions, while avoiding artificial features introduced by detrending. The detectable range is constrained by wavelet resolution, and stable detections are obtained when the signal-to-noise ratio exceeds $\mathrm{SNR} \sim 2$.

Application to GOES soft X-ray observations shows that the method resolves localized time--frequency structures during flare evolution, while no persistent features are found in intervals without oscillations. Differences between Fourier and wavelet representations are consistent with their respective sensitivities to global and time-dependent variability.

These results indicate that combining time--frequency analysis with a time-dependent statistical model provides a useful approach for studying oscillatory signals in non-stationary data.   
    
\section{Data and Code Availability}
\label{data_avail}
The code and data used in this study are publicly available at \cite{yuan_2026_19446731}.
    
\begin{acknowledgments}
The authors acknowledge the use of data from the Geostationary Operational Environmental Satellites (GOES) operated by the National Oceanic and Atmospheric Administration. GOES X-ray flux data were obtained from the NOAA Space Weather Prediction Center, and archived datasets were accessed via the NOAA National Centers for Environmental Information. D.Y. was supported by the National Natural Science Foundation of China (NSFC: 12473050), the Guangdong Natural Science Funds for Distinguished Young Scholars (2023B1515020049), the Shenzhen Science and Technology Project (JCYJ20240813104805008), and the Specialized Research Fund for State Key Laboratory of Solar Activity and Space Weather.
\end{acknowledgments}
   
\bibliography{wavelet_mcmc}
\bibliographystyle{aasjournal}
    
\end{document}